# Nanomechanical sonification of the 2019-nCoV coronavirus spike protein through a materiomusical approach

Markus J. Buehler[a,*]

[a] Laboratory for Atomistic and Molecular Mechanics (LAMM), Massachusetts Institute of Technology, Cambridge, MA, United Stated of America

\* Correspondence: E-mail : mbuehler@MIT.EDU, Phone: +1.617.452.2750

**Abstract:** Proteins are key building blocks of virtually all life, providing the material foundation of spider silk, cells, and hair, but also offering other functions from enzymes to drugs, and pathogens like viruses. Based on a nanomechanical analysis of the structure and motions of atoms and molecules at multiple scales, we report sonified versions of the coronavirus spike protein of the pathogen of COVID-19, 2019-nCoV. The audio signal, created using a novel nanomechanical sonification method, features an overlay of the vibrational signatures of the protein's primary, secondary and higher-order structures. Presenting musical encoding in two versions - one in the amino-acid scale and one based on equal temperament tuning - the method allows for expressing protein structures in audible space, offering novel avenues to represent, analyze and design architectural features across length- and time-scales. We further report a hierarchical frequency spectrum analysis of five distinct protein structures, which offer insights into how genetic mutations, and the binding of the virus spike protein to the human ACE2 cell receptor directly influence the audio. Applications of the approach may include the development of *de novo* antibodies by designing protein sequences that match, through melodic counterpoints, the binding sites in the spike protein. Other applications of audible coding of matter include material design by manipulating sound, detecting mutations, and offering a way to reach out to broader communities to explain the physics of proteins. It also forms a physics-based compositional technique to create new art, referred to as materiomusic, which is akin to finding a new palette of colors for a painter. Here, the nanomechanical structure of matter, reflected in an oscillatory framework, presents a new palette for sound generation, and can complement or support human creativity.

**Audio files:**

- Viral Counterpoint of the Coronavirus Spike Protein (2019-nCoV): https://soundcloud.com/user-275864738/viral-counterpoint-of-the-coronavirus-spike-protein-2019-ncov
- Sonification of the Coronavirus Spike Protein (Amino Acid Scale) https://soundcloud.com/user-275864738/sonification-of-virus-spike

## 1. Introduction

Proteins are the building blocks of virtually all life, providing the material foundation of spider silk, cells, and hair, but also offering other countless functions from enzymes to drugs, and pathogens such as viruses. We cannot see small nanoscopic objects like proteins or other molecules that make up virtually all living matter including our cells, tissues, as well as pathogens such as viruses. Here we apply a computational algorithm to a recently reported protein structure, which allows us to make protein material manifestation audible, by rendering molecular structure information such as vibrational spectra, secondary structure, and folding geometry into a complex audio signal [1–5].

Protein molecules consist of 20 naturally occurring amino acid building blocks that are assembled into hierarchical structures across various length-scales [6–8]. Proteins are usually classified by their structural organization, from primary structure (also known as the sequence of amino acids), secondary structure (local folded geometry: e.g., beta-sheet or alpha-helix), as well as tertiary and quaternary structure (overall folded arrangement). More broadly, the folded structure of proteins, encoded by the sequence of DNA, determines material function and failure in the context of disease or injury [8–14]. This biomaterial organization is reminiscent of the way by which musical compositions can be described, starting from basic harmonic waves (for instance a sine wave), different timbres (overlaying harmonic waves modulated in pitch, volume, and other measures, over time), melodies, chords and multiple chords and melodies played together and against each other. The integrated perspective on all aspects of these assemblies, processed simultaneously through the human auditory system, provides an efficient approach for our brains to understand complex data structures in sound – what humans commonly perceive as "music" [15]. In sonification, we exploit this skill, and use sonification to translate complex data into audible sounds [16,17].

This article focuses on an analysis of the nanomechanical vibrational spectrum of a virus spike protein that is the pathogen of COVID-19, which appears on the surface of the virus as outward facing molecular "spikes", which are critical of the virus' entry into host cells [18]. Figure 1 depicts a rendering of this protein structure, revealing its characteristic spike shape, and also outlines the method used to translate the structure in audible signals based on its vibrational patterning.

The plan of this article is as follows. We begin with a brief review of materials and methods, introducing to approaches used here, and then present specifics of the sonification examples. We discuss the nanomechanical foundation for the approach, and also review basic elements of the musical choices.

**2. Materials and Methods**

Sonification is a method to translate data structures into audible signals. Here, we use a particular approach of sonification termed materiomusic, to use the actual vibrations and structures of molecules to create audible materials that be used for analysis and design. This method has been used in earlier work to translate from material to music and from music to material, due to its reversible and unique mapping [19,20]. The main focus of this paper is on the protein structure with Protein Data Bank (PDB) ID 6vsb. It features three distinct protein chains that are folded into the virus spike protein, consisting of 1,288 residues each (Figure 2 shows an overview of the structure [21]).

**2.1 Sonification of the virus spike protein in amino acid scale**

A protein structure is built up from one-dimensional chains of amino acids. The basis to mapping each of the 20 amino acids onto a unique musical tone is the distinct vibrational spectrum of the molecules as explained in these references: [2,17]. In fact, since each of the amino acids has a unique vibrational spectrum, we may distinguish each of them by its unique sound, or timbre, which is reflected in the overlay of all modal frequencies. Since the eigenmodes are orthogonal, energy does not spill from one mode to another, which implies that they are a fundamental way to understand the vibrational spectrum and sounding of a molecule or assembly. Since each amino acid features a distinct audible expression, defined by their unique profile of molecular vibrations, we can assume that they can be assigned a particular musical "note" or tone. In this framework, the primary sequence of amino acids defines the note sequence played, which is augmented by the other components [19,20]

The next higher level of organization of a protein is the secondary structure, reflecting the local organization of the 1D chain into helices or sheets (or other structures). We incorporate information about the secondary structure associated with each amino acid in the translation step by affecting the duration and volume of notes. During the translation process from protein to music we use the DSSP algorithm to compute the secondary structure from the protein geometry file and sequence [13,22]. As presented in earlier work [19,20] the method of transpositional equivalence is used to scale the frequencies to the audible range, however, while maintaining the accurate ratio of frequencies in the signals generated.

In addition to mapping primary and secondary structure, we realize the overall vibrational motions of the molecule by incorporating their normal mode frequencies into an audio signal, using the method described in [1]. To obtain the normal modes, we use the Anisotropic Network Model [23] approach, and compute the spectra of frequencies. The data is then imported in the Max device [1,24], and sounds are rendered using Ableton Live [25], forming the basis for the secondary signal.

In sound generation, both signals – the primary and secondary, as well as audio generated by the overall vibrations are overlaid and played together, creating a multi-dimensional image of the protein's structure, true to its accurate reflection of vibrational spectra. Sidechain compression is used in some examples to module the volume of the secondary signal based on the level of the primary signal. This renders a dynamic interplay of the two sound sources.

## 2.2 Sonification of the virus spike protein in twelve-tone equal temperament tuning

In the approach described in Section 2.1, all sounds generated are based on the spectrum of the actual molecular vibrations. This, however, poses challenges in matching it with conventional tuning systems. While the sound of amino acids closely resembles that of bells, another approach that can be used is to map each of the 20 amino acids onto a unique tone in a particular musical scale, such as a dorian mode [26,27] as used in the example described in this article. Other possible scale choices could be a major or minor scale, or variations thereof, whereas 20 notes are picked in ascending order to describe each of the 20 amino acids. The note assignments are made in the order TYR, ASN, LEU, MET, GLU, PRO, TRP, ARG, GLN, HIS, PHE, SER, LYS, VAL, ASP, THR, ILE, CYS, ALA, GLY, following an ordering based on their lowest vibrational frequency from small to large.

As in the other approach described in Section 2.1, the next level of organization of a protein is the secondary structure, reflecting the local organization of the 1D chain into helices or sheets, or potentially other structures. We incorporate information about the secondary structure associated with each amino acid in the translation step by affecting the duration and volume of notes, as described in detail in [13,22].

The format used to represent the distance correlation matrix depicted in Figure 3A the concept of rapidly played note insertions, represented musically through strummed notes similar as in a guitar [17]. This approach is further explained in Figure 3, which shows the folding of a protein into a three-dimensional structure. To encode the information that points $i$ and $j$ are neighbors in the folded state, we insert a sequence segment around point $j$ at point $i$. Similarly, a sequence segment around point $i$ is inserted at point $j$, creating a symmetric representation that reflects the geometric fact that points $i$ and $j$ are neighbors. See Figure 3D for the geometry in the folded state, as a visual example of how the coding is implemented in a musical score, in Figure 3E.

Furthermore, additional notes are inserted and played simultaneously, where the volume of the notes is modulated based on the distance of the amino acid inserted from the reference amino acid in the primary sequence, and reflecting the distance correlation matrix shown Figure 3. These reflect the same sequence patterns of the insertions, where a choice can be made about how long the overlaid fragments are. Unlike the insertions described in the previous section, the insertions now added are played on top of the primary notes reflecting the primary and secondary structure of the protein. The resulting complex overlapping melodies represent a kind of counterpoint argument in music – the concept where notes are played against notes – such as used widely by J.S. Bach [28,29]. Counterpoint is a powerful means to manifest references of closeness in a time sequence coded manifestation of signals. It is also the basis for pleasantness of music, whereas a balance of repetition and variation is a foundational structural element. This universality, reminiscent of the universality in proteins represented by secondary structure motifs (such as alpha-helices of beta-sheets), or repetitions of sequence patterns within a protein design – is common across nature.

Importantly, the concept a more generalized counterpoint argument represents physical closeness in musical space in an indirect manner. Similar concepts of counterpunctual approaches are used in many other fields of communication and art, where referencing concepts – albeit distorted – is a powerful way to represent connections in a coded form. For instance, allegories in art can realize connections to multiple touchpoints and associations [30]. These and other concepts are quite similar to the natural structure of proteins, where one residue may be close to many others, as can be directly confirmed in the distance correlation matrix. Moreover, such associations are also reminiscent to other structures, such as the hierarchical organization of neurons in the brain, as visualized in Figure 3B [31].

Through the transformations described above, a total number of 3,647,770 notes are generated in the raw musical coding, reflecting the hierarchical structure of the protein in musical space.

## 2.3 Audio processing

After rendering the sound using Max patches [1,24] and Ableton Live [25], spectral analysis is performance using Sonic Visualizer [32], and the audio files processed using Audacity 2.3.2 [33]. We specifically focus on the melodic spectrogram, which offers a visual representation of how the energy in different frequency bands changes over time. It provides an analytical tool to assess the content of an audio signal, which can complement or validated audible inspection.

## 3. Results

### 3.1 Sonification in amino acid scale

To show the potential application to distinguish protein structures through sound, we depict a spectral comparison of the overall vibrational spectrum of proteins with Protein Data Bank (PDB) [34] identification code 6vsb and 5i08 in Figure 4. Figure 4A shows their normal mode vibrational frequencies, and the Figure 4B a spectrogram analysis of the vibrational signal generated of these proteins. Representative audio signals are included as Supplementary Material (comparison_5i08-6vsb.wav), where the distinction is clearly audible, between 5 distinct protein conformations (the order of tones is as depicted in Figure 4B to 4D, repeated twice overall, representing a total of 10 tones). This example illustrates how the audible system is capable to quickly discern differences in the spectrum, and hence, the vibrational context of a protein structure. This method to assess vibrational changes can be useful to investigate the impacts of sequence changes, or to screen for effects across a very large number of protein data. Figure 4C shows the normal mode vibrational frequencies of three protein complexes – comparing scenarios of the virus spike protein attached (6PDB ID m17) with proteins without the virus spike protein (PDB IDs 6m18 and 6m1d) (for a detailed discussion of these protein confirmations, see [35]). Figure 4D shows the frequency spectrum of the audio signal. It can be clearly recognized how the attachment of the virus spike protein with the human cell receptor changes the frequency spectrum, and it can also be clearly heard. You can hear the transition between the attached state vs. the unattached state between the third and fourth tone, respectively. Notably, when looking at the frequency spectrum, the binding of the virus to the ACE2 receptor leads to a lowering of the frequency spectrum.

Coding for the multiple structural levels, the audio file coronavirus_6vsb_AA-scale_excerpt.wav represents a sonification of the novel coronavirus pathogen, with PDB ID 6vsb. A spectral analysis of the piece is depicted in Figure 5A. The file coronavirus_6vsb_AA-scale-complex_excerpt.wav represents a more complex version of the original score, with added musical ornamentation, expressing creative choices. As can be determined, the raw structure of the original completely algorithmic composition is expanded upon, through added musical experimentation, realizing a more complex exploration of the audible content. The total length of the piece is around 35 minutes.

## 3.2 Sonification in twelve-tone equal temperament tuning

What you hear is a multi-layered algorithmic composition featuring both the vibrational spectrum of the entire protein (expressed in sound and rhythmic elements), the sequence and folding of amino acids that compose the virus spike structure, as well as interwoven melodies - forming counterpoint music - reflecting the complex hierarchical intersecting geometry of the protein. The audio file coronavirus_6vsb_chromatic-scale_excerpt.wav represents a sonification of the novel coronavirus pathogen, with PDB ID 6vsb. A spectral analysis of the entire piece is depicted in Figure 5B.

The techniques used to realize a musical composition include a variety of approaches:

- The primary sequence, all overlaid notes, and all insertions are played as, and rendered as lead sound using physical modeling of a koto (a Japanese harp instrument). The use of physical modeling allows us to directly generate sound using a physical basis, accounting for a true reflection of the way by which vibrating strings generate sound.
- Overlaid notes, coded through the simultaneous excitation of the sounds of amino acid residues in close vicinity, are played through arpeggios. An arpeggio is a way to play the notes of a chord in succession, rather than sounding them simultaneously. This approach enables to create a flowing expression of the complex chord progressions generated through the protein structure, to offer a contrast to the simultaneously played notes.

- A variety of classical instruments are used in the realization of these scores, and artistic choices are executed to vary the relative loudness of one over other instruments.

The realization of the musical score includes a total number of 20 independent musical instruments, including: Koto (several), harp (several), strings, strings played pizzicato, celli, solo violin, clarinet (several), oboe, flutes (several), trumpet, fluegelhorn, and percussion. The total length of the piece is around 1:50.

## 4. Discussion and conclusion

This analysis presented two approaches to sonify a large protein structure, to provide access to structural and potentially functional information through sound. It provides not only access to a truthful reflection of protein structures via sound, but also offers intriguing new ideas for musical composition. When applied in this context, human composers may intersect with the compositions generated by alone, forming a new axis of intersection across scales and species. The focus on proteins is motivated since they are the most abundant material component of all living things: Especially their motion, structure and failure in the context of both normal physiological function and disease is a foundational question that transcends academic disciplines and can be a footing to explore fundamental nanomechanical questions.

We focused on developing a model for the vibrational spectrum of the amino acid building blocks of proteins, an elementary structure from which materials in living systems are built on. This concept is broadly important as at the nanoscale observation, all structures continuously move, reflecting the fact that they are tiny objects excited by thermal energy and set in motion to undergo large deformations as is the case in injury or disease, or due to disruptive mutations. The approach allowed us to extract new musical compositions of the coronavirus spike protein as one way to represent nature's concept of hierarchy as a paradigm to create function from universal building blocks, and exploring the vibrational complexity of a large protein design. This sort of bio-inspired music is an expansion of bio-inspired materials design, and can offer new design ideas for geometric and structural patterning.

Indeed, the translation from various hierarchical systems into one another poses a paradigm to understand the emergence of properties in materials, sound, and related systems. The use of sonification and comparisons of music generated by proteins – as a core physical material building block of life – with conventional compositions may shed light into the unique features of the effects of music and sound and interactions with matter and life, including the brain and other non-human forms of life, as demonstrated in earlier work of sonifications of spider webs [36].

The music and its pleasing nature, especially the realization in the dorian mode (coronavirus_6vsb_chromatic-scale_excerpt.wav), is a metaphor for this nature of the virus to deceive the host and exploit it for its own multiplication.  A virus' genome hijacks the host cell's protein manufacturing machinery, and forces it to replicate the viral genome and produce viral proteins to make new viruses from it.  In light of this nature of viruses, the work explores the fine line between the emergent beauty of life and death as opposite poles, an aspect also critically important for nanomaterials and human health. As one listens to the protein we find that the intricate design results in interesting and actually pleasing, relaxing sounds. The reason for this is that the virus spike protein represents a complex, beautiful design (see also Figure 1). Of course, the intricate protein principles at the basis of the virus do not at all convey the deadly impacts this particularly protein is having on the world. From another vantage point, the pleasing aspect of the music shows the deceiving nature of the virus when it enters the cells by binding to exterior receptors, in order to hijacks our body to replicate. The pleasant sounds represent an analogy for the nature of the virus to deceive the host and exploit it for its own multiplication, emerging as the residue after the transactional infections that diminish or kill the host.

A related important aspect of materiomusical coding is the agency of singular events – in which case, albeit a building block in a system may be small, its impact on the greater context of the overall system behavior is outsized. As in many musical compositions, a single note or sound can be a focal point of a composition (such as a accidentals, or out-of-scale notes or chords in music). In materials science, a pair of atoms may be subject to singular forces near a crack tip [37]. In many systems, a complex structure of a composition of building blocks may solely exist for the purpose of a single, unique, singular note, one of the significant deep foundational questions of many forms of art. Specifically in the context of music, when played alone that note would not have particular exposition. It is only in the context of the whole piece that it stands out. In proteins, we can see a similarity to this singular perspective in point mutations, where certain changes – such as point mutations – are particularly relevant.

Applications of the approach could include the development of antibodies by designing protein sequences that match, through melodic counterpoints, the binding sites in the spike protein, to identify sequences that create strong adhesion for the interruption of disease progression through protein-based drugs. The novel coronavirus is believed to enter human cells via the ACE2 receptor on the exterior of cells [18,35]. ACE2 is an enzyme attached to the outer surface of cells found in our lungs, heart, arteries, kidney, and intestines (whereas one of its natural role is blood pressure regulation). One possible engineering task could be to create a protein molecule, through counterpoint composition, that binds more easily to the virus spike, hence shutting down the pathway of infection. An analysis of the mechanism of interaction between ACE2 and the spike protein can perhaps provide useful insights, and the intersection of these two proteins could be subject of future experimentation in musical space. Not only could this be tackled using human writing, but could also be performed using AI models, as demonstrated in earlier work [2,20]. The data depicted in Figure 4C clearly demonstrated that the binding of the virus spike protein to the human cell leads to an audible change in the sounds generated.

Other aspects by which vibrations of the virus proteins play an important role include our understanding of the virus stability as temperatures increase, for example. Vibrational spectra may also tell us why the SARS-CoV-2 spike gravitates toward human cells more than other viruses. Further applications of materiomusical coding of matter include *de novo* material design by manipulating sound, detecting mutations, and offering a way to reach out to broader communities to explain the science of nanomaterials such as proteins. It also forms a physics-based compositional technique to create new art, referred to as materiomusic, which can be explained akin to finding a new palette of colors for a painter. Here, the nanomechanical structure of matter, reflected in an oscillatory framework, presents a new palette for sound generation. A summary of the new palette of sounds, specifically the 20 amino acid tones, are summarized graphically in Figure 6. The bio-nano-science of molecular motions can hence offer several avenues of future investigation and applications in the interrelated fields of science, art and engineering.

**Acknowledgements:** This work was supported by MIT CAST via a grant from the Mellon Foundation, with additional support from ONR (grant # N00014-16-1-2333) and NIH U01 EB014976.

**Supplementary Files:**

- **comparison_5i08-6vsb.wav:** Comparison of the overall molecular frequency spectrum of two distinct spike proteins from two different coronaviruses, calculated by a normal mode analysis. The analysis of the spectral content is shown in Figure 4. https://soundcloud.com/user-275864738/comparison-5i08-6vsb
- **coronavirus_6vsb_AA-scale_excerpt.wav:** Sonification of the novel coronavirus pathogen, with PDB ID 6vsb (excerpt).
- **coronavirus_6vsb_AA-scale-complex_excerpt.wav**: A more complex version of the original score in amino acid scale, with added musical ornamentation, expressing creative choices (excerpt) https://soundcloud.com/user-275864738/sonification-of-virus-spike
- **coronavirus_6vsb_chromatic-scale_excerpt.wav:** Sonification of the novel coronavirus pathogen, with PDB ID 6vsb using twelve-tone equal temperament and a dorian mode, and realized in a classical orchestral setting (excerpt). https://soundcloud.com/user-275864738/viral-counterpoint-of-the-coronavirus-spike-protein-2019-ncov

Other audio files of proteins and other sonified materials phenomena can be accessed at: https://soundcloud.com/user-275864738

**Figures and captions**

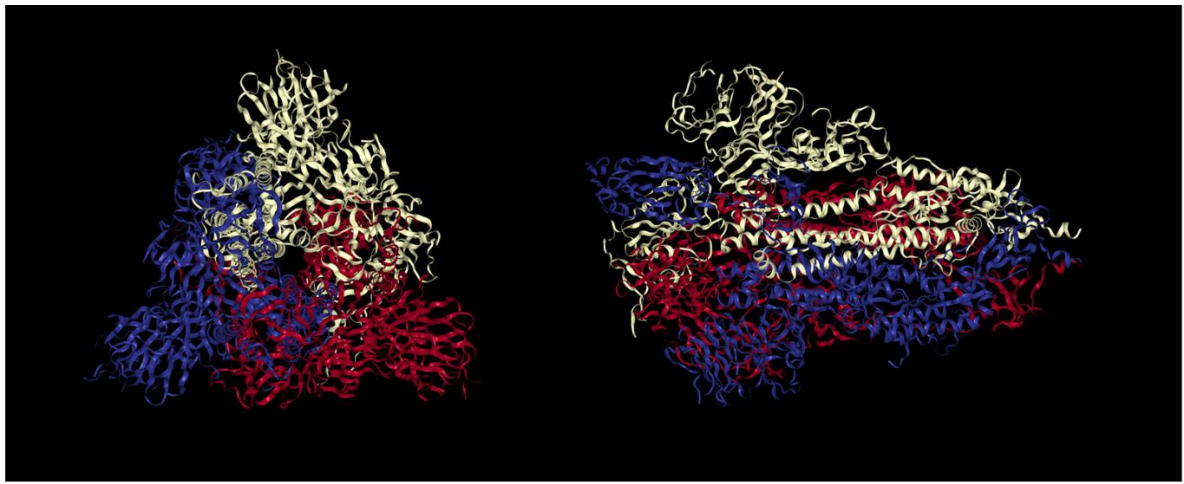

| | | |
|---|---|---|
| Amino acid sequences arranged in space | 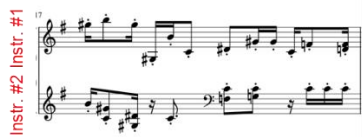 | Combination of melodies played by distinct instruments into chords, interacting melodies, etc.) |
| Amino acid sequence and secondary structure | 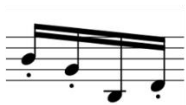 | Melody and rhythm (combination of distinct notes in a pattern) |
| Amino acid (each with unique frequency spectrum) | 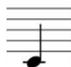 | Note played at different pitch or reflecting different sounds, e.g. distinct sounds generated by amino acids (can also be selected from set of discrete pitches in a musical scale) |
| Interaction of normal modes with environment (e.g. excitation of certain modes - equipartition) | 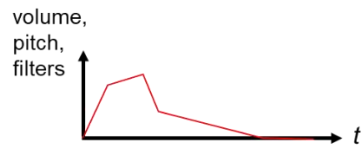 | Time modulation of sine wave or sets of sine waves e.g., volume, pitch, filters |
| Normal mode of chemical structure (fundamental vibrations) | 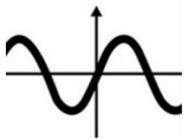 | Sine wave (fundamental vibrations) |

**Figure 1:** Rendering of the spike protein of the novel coronavirus COVID-19, where the three amino acid chains are depicted in distinct colors. Top: The left image shows a top view, the right image a side view. The spike protein is the structure that sticks out of the virus and provides contact with the host, and as such, is an integral part both to understand infection and treatment options. Bottom: Review of the approach used to translate protein structures into sound, by using a physics based method that stems from the analysis of normal modes in the determination of sound [1,2,17].

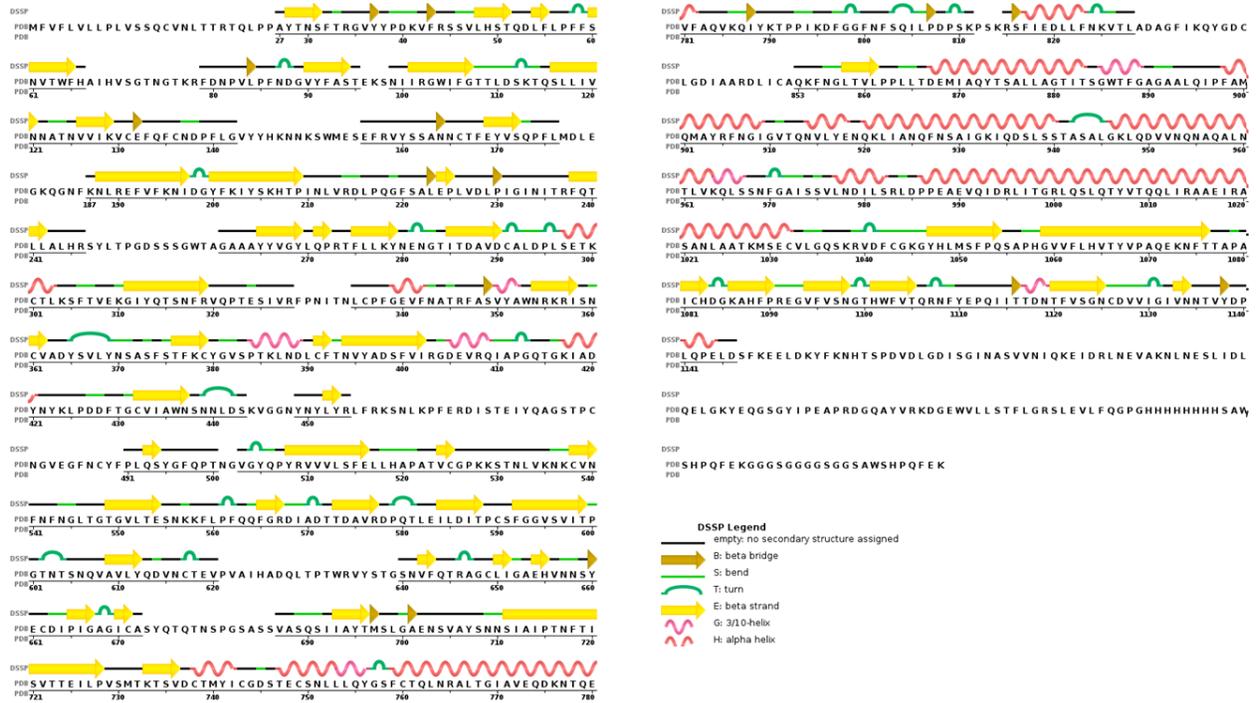

**Figure 2:** Sequence Chain View of one of the three chains of the coronavirus spike protein 6vsb, extracted from the Protein Data Bank [38], and visualized using a sequence analysis tool [21]. The analysis shows that the protein features 16% helical and 21% beta sheet content. The image provides a direct visualization of the various secondary structure motifs that make up the protein, which in turn fold into complex shapes. All this information is incorporated into the musical expression of the protein.

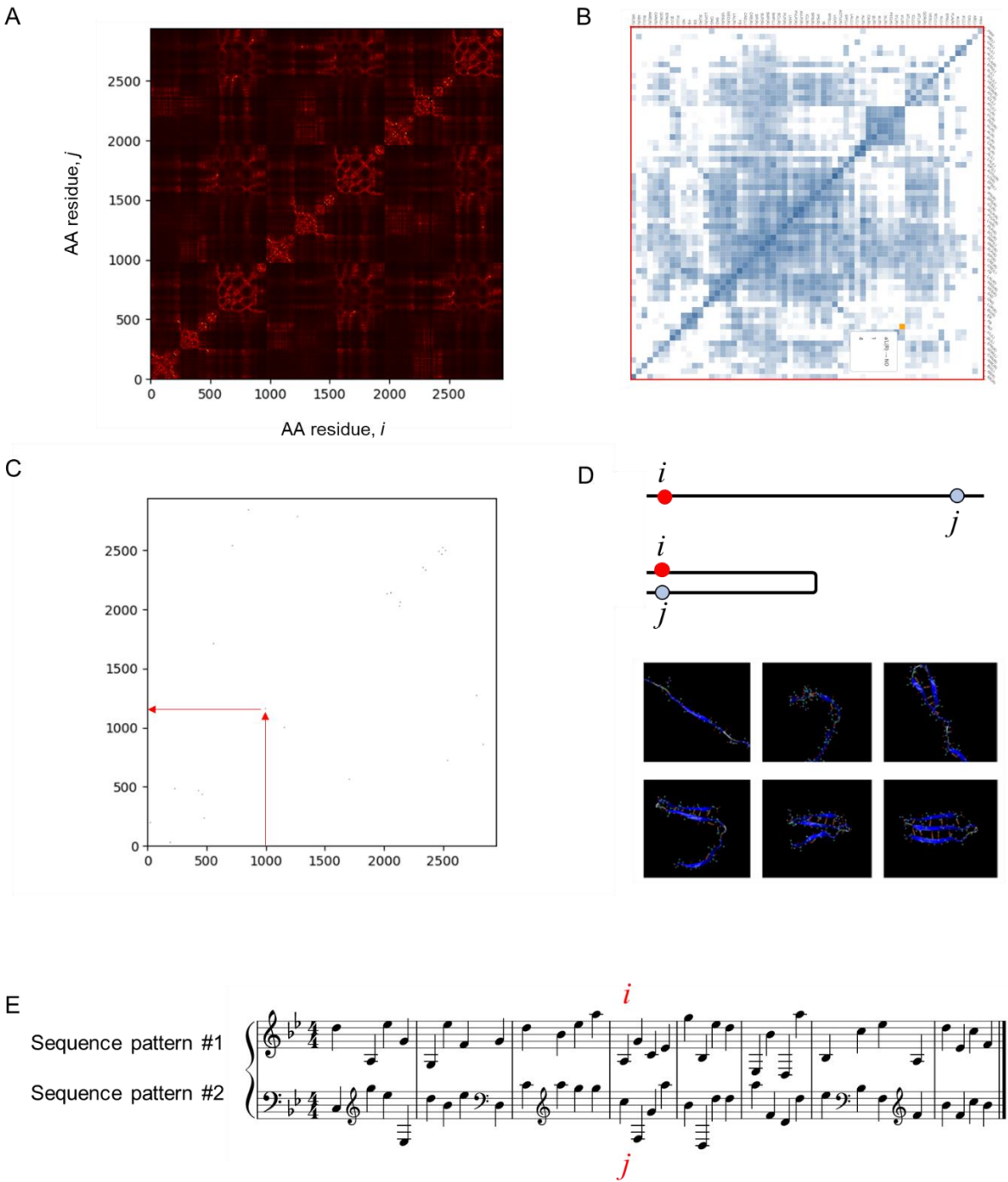

**Figure 3:** Distance correlation matrix for the virus spike protein with PDB ID 6vsb. The plot in panel A shows amino acid residues that are close to each other in brighter red colors, providing an overall image of the protein's 3D folded structure. Panel B depicts the connectome as reported in [31], and rendered via the neuPRINT server [39], showing the similarity of folding concepts in the neural network design in the brain. Panel C shows a plot with only the closest neighbors highlighted, with distances less than 4.5A. Panel D depicts, schematically, how the overall folding geometry is coded in that way. The inlay depicts the folding dynamics of a beta-sheet protein from an initial elongated unstructured protein. Panel E shows an example how two sequence patterns are assembled in the overlapped form, creating note-against-note musical content.

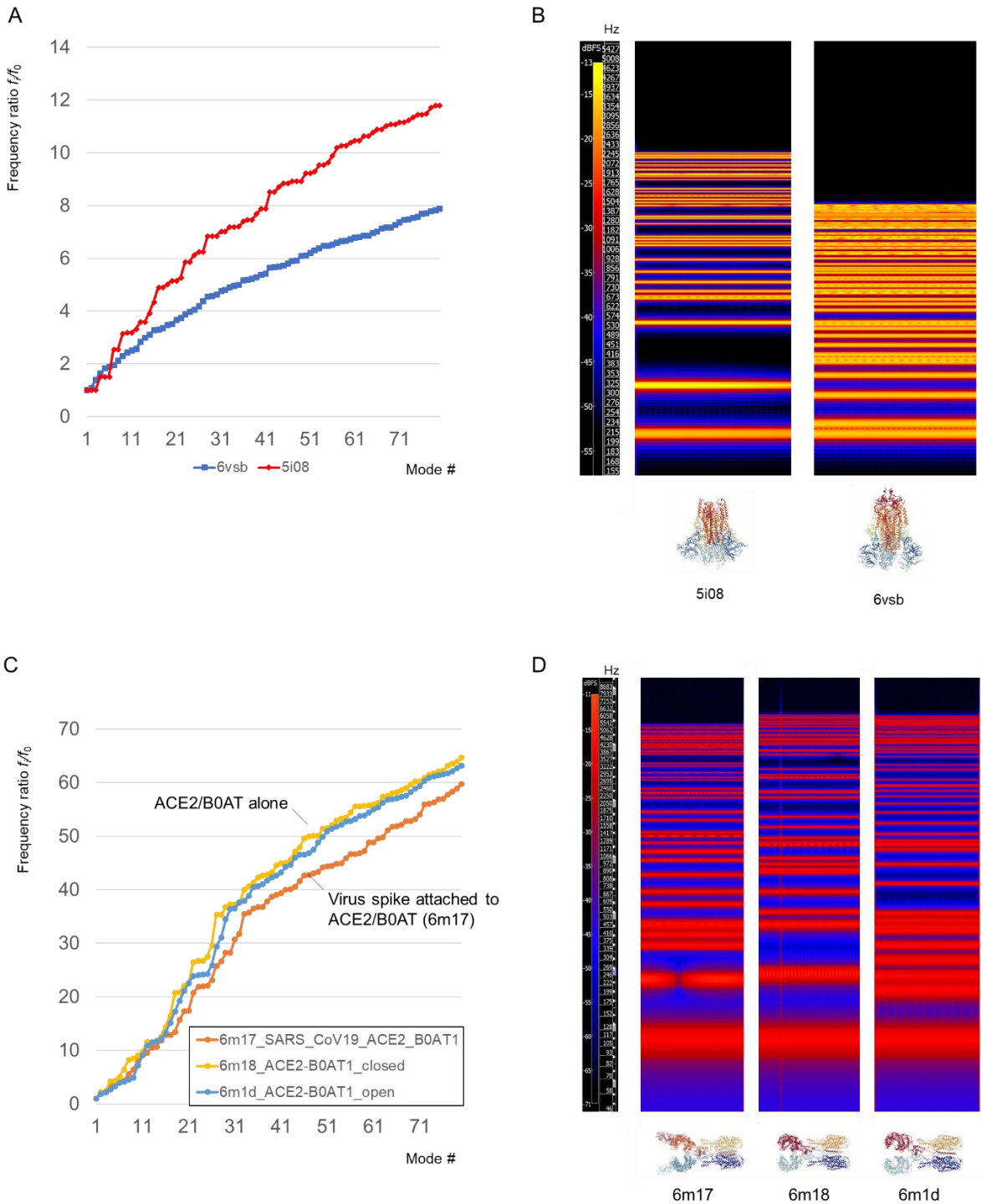

**Figure 4:** Comparison of the vibrational spectrum of two virus spike proteins, here showing 6vsb (right, pathogen of the novel coronavirus COVID-19 that emerged in late 2019) and 5i08 (left, pathogen of human betacoronavirus that causes mild yet prevalent respiratory disease [40]). Panel A shows their normal mode vibrational frequencies, and the panel B a spectrogram analysis of the vibrational signal generated of these proteins. Panel C shows the normal mode vibrational frequencies of three protein complexes – comparing scenarios of the virus spike protein attached (6m17) with proteins without the virus spike protein (6m18 and 6m1d). Panel D shows the frequency spectrum of the audio signal. The audio file comparison_5i08-6vsb.wav provides a direct comparison of the five signals (5i08, 6vsb [novel coronavirus], 6m17 [virus spike attached to human cell], 6m18 and 6m1d [virus not attached to human cell]).

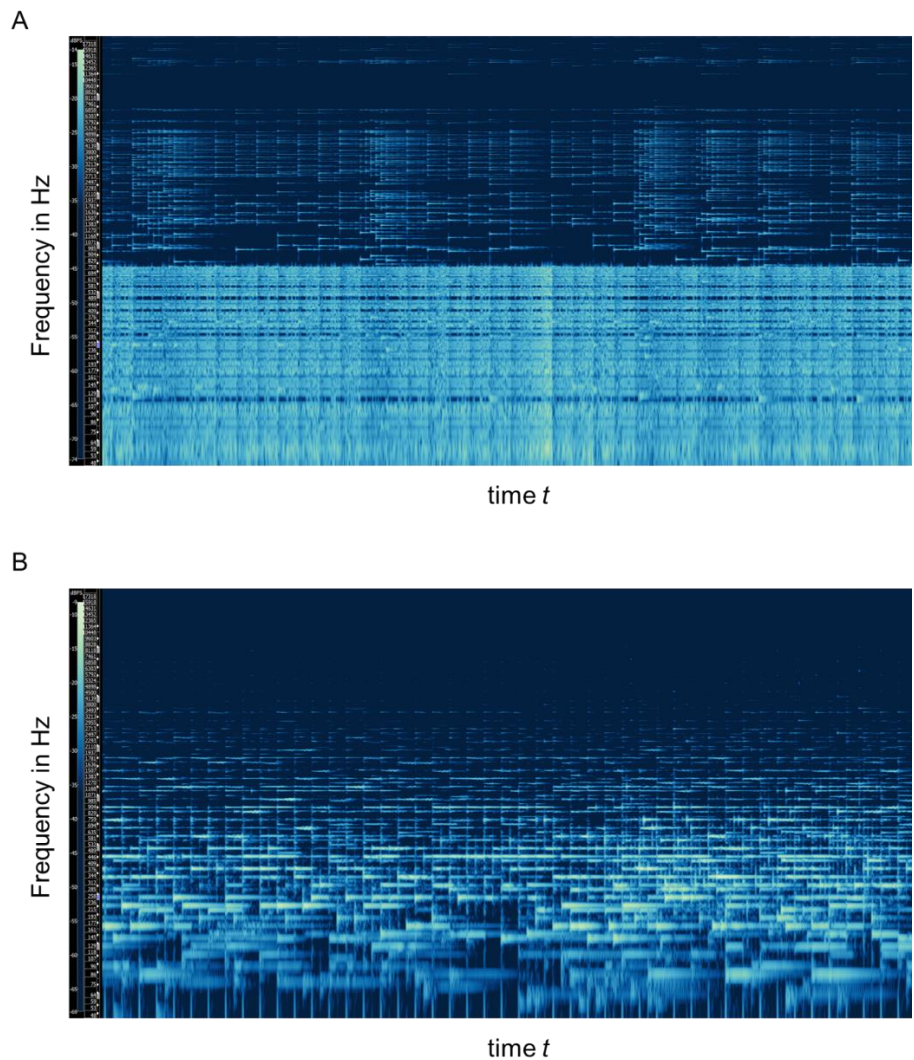

**Figure 5:** Spectral analysis of two musical representation of the coronavirus spike protein with PDB ID 6vsb. Panel A: Realization in amino acid scale, and Panel B: Realization in a chromatic scale, with additional musical ornamentation. The peculiar distribution of vibrational energy in panel, with significant content at low frequencies, is due to the background signal of the overall molecular vibrations that is played together with the amino acid sequence melody.

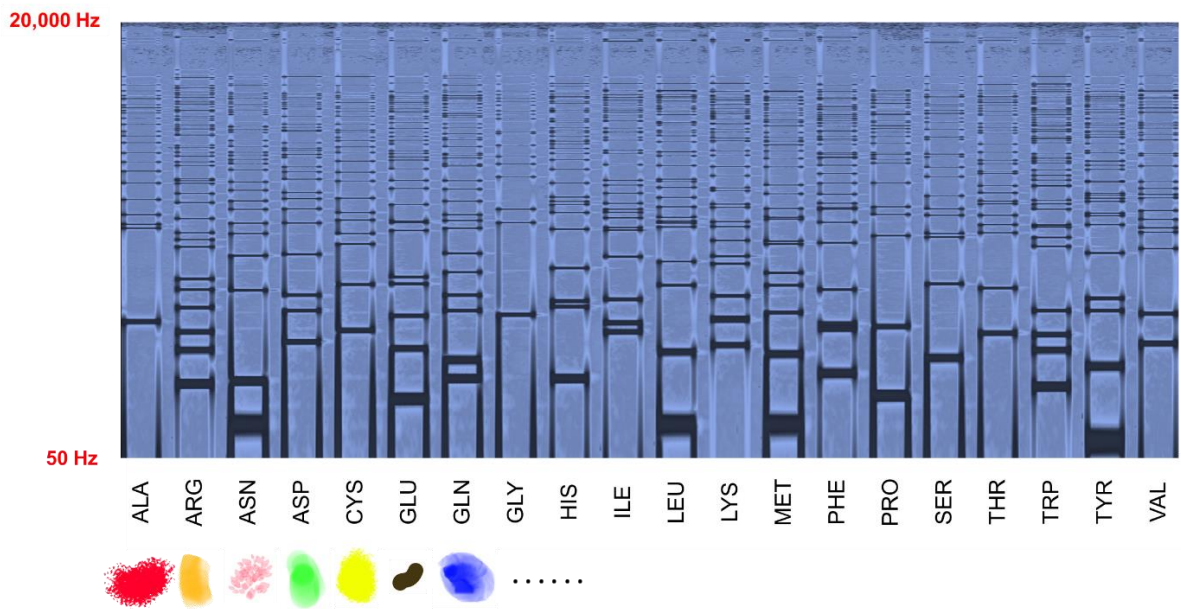

**Figure 6:** Adapted from [19], a visual representing of the sonified barcodes of each of the 20 amino acids. Each of these notes can be viewed as a novel type of building blocks that can be used to generate new art, similar to new colors or new paint materials or paint strokes (see bottom row for examples). The construction of musical art within the constraint of these sets of vibrations offers an interesting challenge in the design of novel music, as exemplified in the examples reported in this article. An example of a multi-protein orchestration, see https://soundcloud.com/user-275864738/orchestra-of-amino-acids.